\documentclass[11pt]{article}
\usepackage[left=2.5cm,top=2.5cm,right=2cm,bottom=2.5cm]{geometry}
\usepackage{amsmath, amssymb}

\usepackage{graphicx,subfigure,epsfig,epsf,color}
\usepackage[utf8x]{inputenc}
\usepackage{slashed}
\usepackage{feynmf}
\usepackage{graphicx}
\usepackage{graphics}
\usepackage{epstopdf}
\usepackage{subfigure}
\usepackage{amsfonts}
\usepackage{sectsty}
\usepackage{hyperref}
\usepackage{cite}
\usepackage{pdflscape}
\usepackage{lipsum}
\begin{document}
 \date{}

\title{Inflationary Cosmology in the Modified $f(R, T)$ Gravity }
\maketitle
 \begin{center}
\author{Ashmita}\footnote{p20190008@goa.bits-pilani.ac.in}
~Payel Sarkar\footnote{p20170444@goa.bits-pilani.ac.in}~Prasanta Kumar Das\footnote{Corresponding author: pdas@goa.bits-pilani.ac.in} \\
 \end{center}
 
 \begin{center}
 Birla Institute of Technology and Science-Pilani, K. K. Birla Goa Campus, NH-17B, Zuarinagar, Goa-403726, India
 \end{center}
 \vspace*{0.25in}

\begin{abstract}
In this work, we study the inflationary cosmology in modified  gravity theory $f(R, T) = R + 2 \lambda T$ ($\lambda$ is the modified gravity parameter) with three distinct class of inflation potentials (i) $\phi^p e^{-\alpha\phi}$, (ii) $(1-\phi^p)e^{-\alpha\phi}$ and (iii) $\frac{\alpha\phi^2}{1+\alpha\phi^2}$ where $\alpha$, $p$ are the potential  parameters. We have derived the Einstein equation, potential slow-roll parameters, the scalar spectral index $n_s$, tensor to scalar ratio $r$ and tensor spectral index $n_T$ in modified gravity theory. We obtain the range of $\lambda$ using the spectral index constraints in the parameter space of the potentials. Comparing our results with PLANCK 2018 data and WMAP data, we found out the modified gravity parameter $\lambda$ lies between $-0.37<\lambda<1.483$.  
\end{abstract}
\section{Introduction}
A number of recent observational findings indicate that we are living in an accelerating universe \cite{Starobinsky, Guth, Linde}, such results are from redshift of type Ia supernova\cite{Reiss}, Cosmic Microwave Background (CMB) \cite{Kolb, Spergel} anisotrpy from Planck\cite{PLANCK}, Wilkinson Microwave Anisotropy Probe (WMAP)\cite{WMAP}, Baryon Acoustic Oscillations (BAO)\cite{Anderson}, Large Scale Structures\cite{spergel1}. This cosmic acceleration can be explained using two different approaches, one by introducing the Cosmological constant model i.e  $\Lambda$CDM\cite{Sahni, Ratra, caroll,Turner} which has negative pressure\cite{Arturo} and the other is by modification of usual Einstein's gravity.\\ Although the idea of Cosmological constant is simpler to describe inflation but it faces few problems such as fine tuning problem\cite{Sahni, Weinberg}, coincidence problem\cite{Sahni, Zlatev}, etc.
\noindent Though the classical theory of general relativity is unquestionably the most suited model of gravity but still it does not fit with the cosmological data which has been regarded as one of the primary drivers behind research into alternate theories of gravity.\\
In this approach, Einstein-Hilbert action is modified by adding some polynomial function of Ricci scalar $R$ (i.e. $f(R)$ gravity)\cite{Nojiri,samanta, buchdahl,capozziello,clifton}, or some function of the Ricci scalar $R$ and/or the trace of the energy-momentum tensor $T$ ($f(R,T)$ gravity) \cite{Harko,Myrzakulov} or some Gauss-Bonnet function ($f(G)$ gravity) \cite{Nojiri2} etc. Among these, $f(R, T)$ gravity has gained popularity recently since it can be used to explain a variety of astrophysical problems such as Inflation\cite{Bhattacharjee}, Dark energy \cite{Bhatti}, Dark matter \cite{Zaregonbadi}, Wormhole \cite{Moraes2}, Gravitational waves etc \cite{Gamonal, sahoo, Sahoo2, Goncalves}.\\
 %%%%%%%%%%%%%%%%%%%%%%%%%%%%%%%%%%%%%%%%%%%%%
 The most straightforward method to study inflation is to consider a scalar field called Inflaton, which under the influence of a particular potential along with the slow-roll approximation (where the kinetic terms are neglected with respect to the potential term) is used in order to examine the inflationary scenario \cite{Baumann,Kinney}. This type of inflaton potential in modified gravity has been extensively studied in many literatures \cite{Alberto, Martin2, Chowdhury, Biswajit,Martin3} along with various cosmological parameters, density perturbation, power-spectrum has been verified by CMB anisotropy measurement\cite{Martin,PLANCK}. The first work of Inflation in modified gravity was done in \cite{Bhattacharjee} using a quadratic potential. Same type of analysis has been shown using non-minimal power-law potential, natural and hill-top potentials in modified gravity\cite{Gamonal}. Starobinsky type potential can also predict compatible results with observational data in $f(R,T)$ gravity.\\
 In this manuscript, we have studied some aspects of inflationary cosmology in a modified $f(R,T)$ gravity theory using a class of three distinct inflaton potentials. Working under the slow-roll approximation, we obtain limits on the modified gravity parameter $\lambda$   and determine the values CMB spectral index parameters which match with the data given by PLANCK2018 and WMAP. \\
The paper is organised as follows: In section 2, we obtain the Einstein Field equations in $f(R,T)$ gravity and derive the slow-roll parameters in this modified gravity theory. We apply these conclusions to numerous inflationary models using the updated formulas for the slow-roll parameters. In section 3,
the inflationary scenario has been discussed for three different potentials, and the cosmological parameters such as scalar spectral index $n_s$, tensor to scalar ratio $r$, tensor spectral index $n_T$ have been derived. These parameters have been subject to constraints in the parameter space of potential $(\lambda,\alpha,p)$ within the context of modified gravity. In section 4, we analyse our results and compare those  with the PLANCK 2018\cite{PLANCK} and the WMAP\cite{WMAP} data.
%%%%%%%%%%%%%%%%%%%%%%%%%%%%%%%%%%%%%%%%%%%%%%%%%%%%%%%%
\section{Field equations in Modified gravity:}
The Einstein-Hilbert action of the modified $f(R,T)$ theory of gravity, in presence of matter can be written as,
\begin{equation}
    \mathcal{S} = \frac{1}{16 \pi G} \int d^4x \sqrt{-g}~f(R,T)+ \int d^4x\sqrt{-g}~\mathcal{L}_m
    \label{action}
\end{equation}
where $R$ is the trace of the Ricci curvature tensor $R_{\mu \nu}$, $T$ is the trace of energy-momentum tensor, $f(R,T) $ is an arbitrary function of $R$ and $T$, $g$ is the determinant of the metric tensor $g_{\mu\nu}$ and G is the Newtonian constant of Gravitation. \footnote{We have used natural units, $c = \hbar = 1$ and considered $8\pi G=1$}. $\mathcal{L}_m$ is the matter Lagrangian and is related to the energy-momentum tensor $T_{\mu\nu}$ as,
\begin{equation}
    T_{\mu\nu}=-\frac{2}{\sqrt{-g}}~\frac{\delta}{\delta g^{\mu\nu}}(\sqrt{-g}\mathcal{L}_m)
\end{equation}
By taking the metric variation of the action (\ref{action}), we find the modified Einstein equation as
\begin{equation}
    f_R(R,T)~R_{\mu\nu}-\frac{1}{2}f(R,T)~g_{\mu\nu}+(g_{\mu\nu}\Box-\nabla_{\mu}\nabla_{\nu})~f_R(R,T)=T_{\mu\nu}-f_T(R,T)~T_{\mu\nu}-f_T(R,T)~\theta_{\mu\nu}
    \label{field}
\end{equation}
where 
$f_R(R,T)=\frac{\partial f(R,T)}{\partial R}$, $f_T(R,T)=\frac{\partial f(R,T)}{\partial T}$ and $\theta_{\mu\nu}=g^{\alpha\beta}\frac{\partial T_{\alpha\beta}}{\partial g^{\mu\nu}}$.  The energy momentum tensor for perfect fluid is, $T_{\mu\nu}=(\rho+p)u_{\mu}u_{\nu}-pg_{\mu\nu}$, where $u^{\mu}$ is the four velocity of the perfect fluid satisfying $u_{\mu}u^{\mu}=1$ in the comoving frame. 
 We choose the matter Lagrangian such that, $\mathcal{L}_m=-p$ which yields $\theta_{\mu\nu}=-2T_{\mu\nu}-pg_{\mu\nu}$. Here $\rho$ and $p$ are the energy density and pressure respectively,  There are many functional forms of $f(R,T)$ available in different literatures\cite{Harko, Singh, Jamil}. Here we have chosen $f(R,T)=R+2f(T)=R+2\lambda T$, $\lambda$ is a constant. Considering the above form of $f(R,T)$, the field equation given by Eq.~(\ref{field}) gives,
 \begin{equation}
 \label{Einstein}
     R_{\mu\nu} - \frac{1}{2} g_{\mu\nu} R = T_{\mu\nu}^{eff}
 \end{equation} 
 where
 $T_{\mu\nu}^{eff}=T_{\mu\nu}+2\lambda T_{\mu\nu}+2\lambda pg_{\mu\nu}+\lambda Tg_{\mu\nu}$.
 Assuming that the Universe is filled up with a single and homogeneous inflaton field, the effective energy momentum tensor of the inflaton field will take a diagonal form and we can define the effective energy density and pressure as,
 \begin{equation}
 \label{effectiveTmunu}
     T_{00}^{eff}=\rho_{\phi}^{eff}=\frac{1}{2}\dot{\phi}^2(1+2\lambda)+V(\phi)(1+4\lambda),
     ~~ T_{ij}^{eff}=p_{\phi}^{eff} \delta_{ij}=\left(\frac{1}{2}\dot{\phi}^2(1+2\lambda)-V(\phi)(1+4\lambda)\right) \delta_{ij}
 \end{equation}
 Hence, the equation of state parameter $\omega^{eff}_{\phi}$ will be,
 \begin{equation}
     \omega_{\phi}^{eff}=\frac{p_{\phi}^{eff}}{\rho_{\phi}^{eff}}=\frac{\dot{\phi}^2(1+2\lambda)-2V(\phi)(1+4\lambda)}{\dot{\phi}^2(1+2\lambda)+2V(\phi)(1+4\lambda)}
 \end{equation}
\noindent 
The trace of energy momentum tensor can be obtained from Eq.~(\ref{effectiveTmunu}) as,
\begin{equation}
     T^{eff}=\rho_{\phi}^{eff}-3p_{\phi}^{eff}=-\dot{\phi}^2(1+2\lambda)+4V(\phi)(1+4\lambda)
\end{equation}
\noindent  
The line element for the Friedman-Lemaitre-Robertson-Walker (FLRW) metric in spherical coordinates has the following form,
 \begin{equation}
     ds^2=dt^2-a^2(t)\left[\frac{dr^2}{1-K r^2} + r^2 d\theta^2 + r^2 \sin^2 \theta d\phi^2  \right]
     \label{metric}
 \end{equation}
The effective FRW equation in modified $f(R,T)$ gravity can be derived as,
\begin{equation} 
    3H^2=\rho_{\phi}^{eff},
    ~~ 2\dot{H} + 3H^2= - p_{\phi}^{eff}
    \label{hubble}
\end{equation}
In the above equations $a(t)$ is the scale factor, $H=\frac{\dot a}{a}$ is the Hubble parameter and the dot represents the derivative with respect to time($t$). The continuity equation for $\rho_{\phi}^{eff}$ and $p_{\phi}^{eff}$ can be derived as,
\begin{equation}
    \dot{\rho_{\phi}}^{eff} + 3H(\rho_{\phi}^{eff} + p_{\phi}^{eff})=0
\end{equation}
which gives
\begin{equation}
\label{EOM}
    \ddot{\phi}~(1+2\lambda) + 3H\dot{\phi}~(1+2\lambda)+V_{,\phi}~(1+4\lambda)=0
\end{equation}

\subsection{Slow-roll parameters and CMB constraints:}
We assumed that the universe is filled up with a scalar field which is minimally coupled to modified gravity. We intend to employ the slow-roll approximation to different inflaton potential to study the spectral index parameters given by CMB. We can define the first slow-roll parameter as \cite{Gamonal},
\begin{equation}
\label{epsilon}
    \bar{\epsilon}=-\frac{\dot{H}}{H^2}=\frac{3}{2}\left[\frac{\dot{\phi}^2(1+2\lambda)}{\frac{1}{2}\dot{\phi}^2(1+2\lambda)+V(\phi)(1+4\lambda)}\right]
\end{equation}
Under the slow-roll approximation, we find
\begin{equation}
\label{condition}
    \dot{\phi}(1+2\lambda)<<V(1+4\lambda),~~ 3H\dot{\phi}(1+2\lambda)=-V_{,\phi}(1+4\lambda)
\end{equation}\\
Applying these in Eq.~(\ref{epsilon}), we find the slow-roll parameter $\bar{\epsilon}$ for $f(R,T)$ gravity as,
\begin{equation}
\label{epsilonv}
   \bar\epsilon=\frac{3\dot{\phi}^2(1+2\lambda)}{2V(1+4\lambda)}=\frac{1}{2(1+2\lambda)}\left(\frac{V_{,\phi}}{V}\right)^2=\bar{\epsilon}_v
\end{equation}
Similarly by taking the derivative on Eq. (\ref{condition}), we can define the second slow-roll parameter as,
\begin{equation}
\label{etav}
    \bar{\eta}=-\frac{\ddot{\phi}}{H\dot{\phi}}=\frac{1}{1+2\lambda}\left(\frac{V_{,\phi\phi}}{V}\right)= \bar{\eta}_v
\end{equation}
where $\bar\epsilon$ and $\bar\eta$ represents Hubble slow roll parameters whereas $\bar{\epsilon}_v$ and $\bar{\eta}_v$ represents potential slow-roll parameters.
The amount of inflation, required to produce isotropic and homogeneous universe, is described by the e-fold number $N$, which can be derived from Eq.~(\ref{condition}) and Eq.~(\ref{hubble})as,
\begin{equation}
\label{efold}
N=-\int\frac{H}{\dot{\phi}}d\phi=-(1+2\lambda)\int_{\phi_{in}}^{\phi_{final}}\frac{V}{V_{,\phi}}d\phi
\end{equation}
Here, $\phi_{final}$ is calculated by taking $\bar{\epsilon}_v=1$, which corresponds to the end of inflation and $\phi_{in}$ is the initial value of inflation field at the beginning of inflation.
The CMB parameters i.e. scalar spectral index $n_s$, scalar to tensor ratio $r$ and tensor spectral index $n_T$ can be written in terms of slow roll parameters as,
\begin{equation}
\label{spectralindex}
    n_s-1=-6 \bar{\epsilon}_v+2\bar{\eta}_v,
    ~~ r=16\bar{\epsilon}_v, ~~ n_t=-2\bar{\epsilon}_v
\end{equation}

\section{Analysis of different inflationary models in modified gravity}
In this section, we calculate the CMBR spectral index parameters for three different inflaton potentials and obtain constraint on the modified gravity parameter $\lambda$ in potential parameter space \cite{sarkar}.
%%%%%%%%%%%%%%%%%%
\subsection{Case 1: Inflaton potential $V=V_0\phi^p e^{-\alpha\phi}$, $\lambda \neq 0$}
To start with, we consider the scalar (inflaton) potential of inflationary expansion as, 
\begin{equation*}
    V(\phi) = V_0 \phi^p e^{-\alpha\phi}
\end{equation*}
where $V_0$ is a constant, $p$ and $\alpha$ are the potential parameters. Under the slow-roll approximation, the potential slow-roll parameters can be obtained (using Eq. (\ref{epsilonv}) and Eq. (\ref{etav})) as,
\begin{equation}
    \bar{\epsilon}_v = \frac{1}{2}\frac{(p- \alpha \phi)^2}{(1+2 \lambda) \phi^2}, ~~
    \bar{\eta}_v = \frac{p^2 + \alpha^2 \phi^2 - p(1+2 \alpha \phi)}{(1+2 \lambda) \phi^2}
\end{equation}
From Eq. (\ref{spectralindex}), the CMBR spectral index parameters $n_s$, $r$ and $n_T$ can be evaluated as follows,
%%%%
\begin{equation}
    n_s = \frac{-p^2 +(1-\alpha^2 +2 \lambda)\phi^2 + 2p(-1+\alpha \phi)}{(1+2 \lambda) \phi^2},~~ r = \frac{8 (p- \alpha \phi)^2}{(1+2 \lambda) \phi^2}, ~~
    n_T = -\frac{ (p- \alpha \phi)^2}{(1+2 \lambda) \phi^2} 
\end{equation}
\noindent We now explore the parameter space $(p, \alpha)$ of the inflaton potential in modified gravity approach - which can produce the desired number of e-fold and estimate the spectral index parameters of CMB. 
%**************************************
\begin{table}[htb]
 \centering
 \addtolength{\tabcolsep}{-0.5pt}
 \small
 \begin{tabular}{|c c c c c c c c c|}
 \hline
 Potential, & $V=V_0\phi^{p}e^{-\alpha\phi}$, & $p = 2$ & & & & & & \\[-0.01ex]
 \hline
    Range of $\lambda$ & $\lambda$ & $\alpha$ & $\phi $ &  $\phi_f$ & N & $n_s$ & r & $n_T$ \\
  $ -0.21580 < \lambda < 0.00783 $ & -0.11410 & 0.01 & 16.51 & 1.59691 & 55 & 0.964983 & 0.128031 & -0.01600 \\
 \hline
  $ -0.08205 < \lambda < 0.24490 $ & 0.06919 & 0.05 & 10.98 & 1.24309 & 56 & 0.964927 & 0.04743 & -0.00593 \\
%  \hline
  $ -0.06555 < \lambda < 0.27860 $ & 0.09024 & 0.1 & 12 & 1.26060 & 53 & 0.964939 & 0.092241 & -0.01153 \\
  \hline\hline
   Potential, & $V=V_0\phi^{p}e^{-\alpha\phi}$, & $p = 4$ & & & & & & \\[-0.01ex]
 \hline
    Range of $\lambda$ & $\lambda$ & $\alpha$ & $\phi $ &  $\phi_f$ & N & $n_s$ & r & $n_T$ \\
  $ 0.21590 < \lambda < 0.15320 $ & 0.18000 & 0.01 & 19.02 & 2.41074 & 63 & 0.95424 & 0.23601 & -0.02950 \\
 \hline
  $ -0.33410 < \lambda < -0.27390 $ & -0.31670 & 0.05 & 30 & 4.41369 & 55 & 0.956810 & 0.15154 & -0.01894 \\
%  \hline
  $ -0.08268 < \lambda < 0.06080 $ & 0.02000 & 0.1 & 18.01 & 2.59366 & 60 & 0.961950 & 0.11468 & -0.01433 \\
  \hline
 \end{tabular}
 \caption{\label{table:1} For $V=V_0 \phi^{p} e^{-\lambda\phi}$, the e-fold number $N$ and the spectral index parameters $n_s$, $r$ and $n_T$ are calculated for a fixed value of $\phi$ and $\lambda$ taken from given range.}
 \end{table}
 %%%%%%%%%%%%%%%%%%%%%%%%%%%%%%%%%%%%%%%%%%%
 In Table (\ref{table:1}), we have shown the values of $n_s$, $r$, $n_T$ respectively for a particular value of scalar field $\phi$ and modified gravity parameter $\lambda$. The range of $\lambda$
is obtained from the $\pm 3 \sigma$ constraints of $n_s=0.9649 \pm 0.0042$ for a  fixed $\phi$. For a particular $\lambda$ value (chosen from the range), and potential parameters ($\alpha, p$), the e-fold and spectral index parameters are calculated which are shown in the Table (\ref{table:1}). Note that, $\phi_f$ is evaluated by taking $\epsilon_v=1$ (exit of inflation) for different values of potential parameters $p=2$, $4$ and $\alpha=0.01$, $0.05$, $0.1$. The range of $\lambda$ mentioned in the table for potential parameters choice (e.g. $-0.21580 < \lambda < 0.00783$ corresponding to $p=2$, $\alpha = 0.01$) can produce the e-fold number $N$ lying between $40 - 70$. Note that for a given $p=2$ (say), as $\alpha$ changes from $0.01$to $0.1$,  $r$ changes from $0.128$ to $0.092$ and $n_T$ changes from $-0.0160$ to $-0.0115$. The spectral index parameters $r$ and $n_T$ estimated above in the potential parameters space can be compared with the existing experimental (PLANCK+BAO) upper bound. From Table.~(\ref{table:1}) we can conclude that for $p=2$ and $\alpha=0.05$, $0.1$ we obtain $n_s$
values lies within $\pm3\sigma$ limit of PLANCK 2018 data along with $r<0.106$ (PLANCK+BAO). For the rest of the choices of $p$ and $\alpha$ although $n_s$ matches with PLANCK2018 data, but $r$ is beyond the limit given by PLANCK+BAO.

\subsection{Case 2: Inflaton potential $V=V_0(1-\phi^p)e^{-\alpha\phi}$, $\lambda \neq 0$}
We next consider the potential of the form   
\begin{equation*}
    V=V_0 (1 - \phi^p) e^{-\alpha\phi}
\end{equation*}
where $p$ and $\alpha$ are the potential parameters. Similarly, the potential slow-roll parameters can be calculated as
\begin{equation}
    \bar{\epsilon}_v = \frac{\bigl\{ p \phi^p -\alpha \phi ~(-1+\phi^p) \bigr\}^2}{2 ~(1+2\lambda) \phi^2 (-1+\phi^p)},~~
%\end{equation}
%
%\begin{equation}
   \bar{\eta}_v = \frac{(-1+p) ~p ~\phi^p -2 p ~\alpha ~\phi^{1+p} + \alpha^2 \phi^2 (-1+\phi^p) }{(1+2\lambda)~\phi^2~(-1+\phi^p)}
\end{equation}
and the CMBR spectral index parameters $n_s$, $r$ and $n_T$ as,

\begin{equation}
    n_s = 1 - \frac{3 \bigl\{ p \phi^p -\alpha \phi~(-1 + \phi^p) \bigr\}^2}{(1+2 \lambda)~ \phi^2 ~(-1 + \phi^p)^2} + \frac{2 \bigl\{ (-1+p) p~\phi^p -2 p~ \alpha \phi^{1+p} + \alpha^2 \phi^2 (-1+\phi^p) \bigr\}}{(1+2 \lambda)~ \phi^2 ~(-1 + \phi^p)^2}
\end{equation}

\begin{equation}
    r = \frac{8 \bigl\{ p \phi^p -\alpha \phi~(-1 + \phi^p) \bigr\}^2}{(1+2 \lambda)~ \phi^2 ~(-1 + \phi^p)^2}, ~~
    n_T = -\frac{ \bigl\{ p \phi^p -\alpha \phi~(-1 + \phi^p) \bigr\}^2}{(1+2 \lambda)~ \phi^2 ~(-1 + \phi^p)^2}
\end{equation}
The power law $\phi^p$ or $1-\phi^p$ type potential do not give good results for the cosmological parameters. The tensor to scalar ratio is fairly high($\sim 0.4$) in comparison to  the PLANCK2018 and WMAP data which gives motivation to choose the combined potentials of power-law and exponent type. Also, in $1-\phi^p$ potential, for $p=2$, the e-fold number blows. It is also one of the main reason to choose these type of combined potentials.\\
The results for this case is displayed in Table (\ref{table:2}).
\begin{table}[htb]
 \centering
 \addtolength{\tabcolsep}{-1pt}
 \small
 \begin{tabular}{|c c c c c c c c c|}
 \hline
 Potential, & $V=V_0(1 - \phi^{p})e^{-\alpha\phi}$, & $p = 2$ & & & & & & \\[-0.01ex]
 \hline
    Range of $\lambda$ & $\lambda$ & $\alpha$ & $\phi $ &  $\phi_f$ & N & $n_s$ & r & $n_T$ \\
  $ -0.37000 < \lambda < -0.26770 $ & -0.32440 & 0.01 & 24 & 2.72518 & 54 & 0.964750 & 0.122985 & -0.01537 \\
 \hline
 $ -0.23480 < \lambda < -0.03140 $ & -0.14180 & 0.05 & 14.99 & 2.08376 & 53 & 0.964964 & 0.078829 & -0.009854 \\
 % \hline
  $ 0.04388 < \lambda < 0.64820 $ & 0.23710 & 0.1 & 10 & 1.69176 & 53 & 0.96498 & 0.05648 & -0.00706 \\
  \hline\hline
   Potential, & $V=V_0(1 - \phi^{p})e^{-\alpha\phi}$, & $p = 4$ & & & & & & \\[-0.01ex]
 \hline
    Range of $\lambda$ & $\lambda$ & $\alpha$ & $\phi $ &  $\phi_f$ & N & $n_s$ & r & $n_T$ \\
  $ 0.42660 < \lambda < 1.48300 $ & 0.50000 & 0.01 & 16 & 2.09830 & 64 & 0.95550 & 0.23041 & -0.02880 \\
 \hline
  $ 0.07318 < \lambda < 0.71010 $ & 0.20000 & 0.05 & 18 & 2.39400 & 65 & 0.96118 & 0.16900 & -0.02118 \\
%  \hline
  $ -0.18670 < \lambda < 0.16430 $ & -0.08000 & 0.1 & 20 & 2.90578 & 63 & 0.96420 & 0.09520 & -0.01190 \\
  \hline
 \end{tabular}
 \caption{\label{table:2} For $V=V_0 (1 - \phi^{p}) e^{-\alpha\phi}$, the e-fold number $N$ and the spectral index parameters $n_s$, $r$ and $n_T$, calculated for a fixed value of $\phi$ and $\lambda$ are presented.}
 \end{table}
 %%%%%
 We see that for $p=2, \alpha=0.1,0.05 $, $n_s$ lies within $\pm 3 \sigma$ of PLANCK2018 data along with $r$ range given by PLANCK+BAO except for $\alpha=0.01$. On the other hand, for $p=4,\alpha=0.1$ the observational parameters values match with experimental data. We also find that for $p=2(4)$, as $\alpha$ changes from $0.01$ to $0.1$, $r$ decreases from $0.12298$ to $0.05648$, $n_T$ changes from $-0.01537$ to $-0.00706$. 
%%%%%%%%%%%%
\subsection{Case 3: Inflaton potential $V=V_0\frac{\alpha\phi^2}{1+\alpha\phi^2}$, $\lambda \neq 0$}
Finally, we consider the potential 
\begin{equation*}
    V=V_0\frac{\alpha\phi^2}{1+\alpha\phi^2}
\end{equation*}
where $\alpha$ is the potential parameter. The potential slow-roll parameters are found to be 
\begin{equation}
    \bar{\epsilon}_v = \frac{2}{(1+2 \lambda)~(\phi + \alpha \phi^3)^2}, ~~
    \bar{\eta}_v = \frac{2 - 6\alpha \phi^2}{(1+2 \lambda)~(\phi + \alpha \phi^3)^2}
\end{equation}
and the spectral index parameters are obtained as,
\begin{equation}
    n_s = 1 - \frac{12 \alpha}{(1+2 \lambda)~(1+\alpha \phi^2)^2} - \frac{8}{(1+2 \lambda)~(\phi + \alpha \phi^3)^2}
\end{equation}
%%%
\begin{equation}
    r = \frac{32}{(1+2 \lambda)~(\phi + \alpha \phi^3)^2}, ~~
    n_T = -\frac{4}{(1+2 \lambda)~(\phi + \alpha \phi^3)^2}
\end{equation}
Here, we have explored the desired number of e-fold($N$) and  the CMB parameters for different values of $\alpha$ of the inflaton potential in modified gravity model. In Table (\ref{table:3}), we have shown the e-fold number $N$ and the spectral index parameters $n_s$, $r$, $n_T$ for a fixed $\phi$ and $\lambda$.
%%%%%%%%%%%%%%%%%
\begin{table}[htb]
 \centering
 \addtolength{\tabcolsep}{-1pt}
 \small
 \begin{tabular}{|c c c c c c c c c|}
 \hline
 Potential, & $V=V_0\frac{\alpha\phi^2}{1+\alpha\phi^2}$ & & & & & & &  \\[-0.01ex]
 \hline
    Range of $\lambda$ & $\lambda$ & $\alpha$  & $\phi $ &  $\phi_f$ & N & $n_s$ & r & $n_T$ \\
  $ 0.10680 < \lambda < 0.79830 $ & 0.32930 & 1 & 3.7 & 0.721898 & 44 & 0.96484 & 0.00653 & -0.00082 \\
 \hline
  $ -0.10680 < \lambda < 0.34250 $ & 0.03791 & 2 & 3.5 & 0.694246 & 43 & 0.96480 & 0.003734 & -0.00047 \\
  \hline
 \end{tabular}
 \caption{\label{table:3} For $V=V_0 \frac{\alpha\phi^2}{1+\alpha\phi^2}$, the e-fold number $N$ and the spectral index parameters are calculated for a fixed value of $\phi$ and $\lambda$. }
 \end{table}
 %%%
 For $\alpha = 1(2)$ and $\lambda=0.3293(0.0379)$, we find $r \sim 0.00653(0.00373)$ and $n_T \sim -0.00082(-0.00047)$ together with the number of e-fold $N=44(43)$ lies well within the range $40-60$. Hence for this particular form of potential, all the cosmological parameters exist within the experimental data range.
%%%%%%%%%%%%%%%%%%%%
\subsection{Case 4: $V=V_0 \phi^p e^{-\alpha\phi}$, $V=V_0(1-\phi^p)e^{-\alpha\phi}$, $V=V_0\frac{\alpha\phi^2}{1+\alpha\phi^2}$ with $\lambda = 0$} 
In this section, we analyze the last three inflationary potentials for $\lambda = 0$ which implies $f(R,T) = R + 2 \lambda T = R$, i.e. normal Einstein gravity. 
In Table \ref{table:4}, we have tabulated the values of $N$, $n_s$, $r$, $n_T$ for $p=2$, $4$ and $\alpha=0.01$ and $0.1$ for the potentials $\phi^p e^{-\alpha\phi}$, $(1-\phi^p)e^{-\alpha\phi}$ and $\frac{\alpha\phi^2}{1+\alpha\phi^2}$, respectively. From the table, we see that although the e-fold lies in the range $40-70$ and $n_s$ value matches with the experimental data but $r$ value is little higher than the PLANCK+BAO data for some values of $\alpha$ and $p$.
\begin{table}[htb]
 \centering
 \addtolength{\tabcolsep}{-0.5pt}
 \small
 \begin{tabular}{|c c c c c c c|}
 \hline
 Potential, & $V=V_0\phi^{p}e^{-\alpha\phi}$, & $p = 2$, & $\lambda = 0$ & & & \\[-0.01ex]
     $\alpha$ & $\phi $ &  $\phi_f$ & N & $n_s$ & r & $n_T$ \\
     \hline
     0 & 12.73 & 1.41400 & 40 & 0.95063 & 0.19747 & -0.01653\\ 
     0.01 & 14.48 & 1.40428 & 55 & 0.964507 & 0.131321 & -0.0164151 \\
     0.05 & 13.05 & 1.36592 & 54 & 0.96585 & 0.0852956 & -0.0106619 \\
     0.1 & 11.486 & 1.32082 & 41 & 0.964186 & 0.0439562 & -0.005494 \\
     \hline
      Potential, & $V=V_0 \phi^{p} e^{-\alpha\phi}$, & $p = 4$, & $\lambda = 0$ & & & \\[-0.01ex]
     $\alpha$ & $\phi $ &  $\phi_f$ & N & $n_s$ & r & $n_T$ \\
     \hline
     0 & 18.13 & 2.82800 & 40 & 0.92698 & 0.38942 & -0.0487\\
     0.01 & 25.44 & 2.80857 & 62 & 0.9543 & 0.233916 & -0.029239 \\
     0.05 & 19.755 & 2.73184 & 58 & 0.95625 & 0.186002 & -0.023250 \\
     0.1 & 17.32 & 2.64164 & 53 & 0.956185 & 0.137177 & -0.017147 \\
     \hline
     Potential, & $V=V_0(1-\phi^{p})e^{-\alpha\phi}$, & $p = 2$, & $\lambda = 0$ & & & \\[-0.01ex]
     $\alpha$ & $\phi $ &  $\phi_f$ & N & $n_s$ & r & $n_T$ \\
     \hline
     0.01 & 14.55 & 1.92403 & 54 & 0.964424 & 0.131296 & -0.016412\\
     0.05 & 13 & 1.89392 & 52 & 0.964932  & 0.08780  & -0.010975\\
     0.1 & 11.65 & 1.8588 & 53 & 0.964547 & 0.042571  & -0.005321\\
     \hline
      Potential, & $V=V_0(1-\phi^{p})e^{-\alpha\phi}$, & $p = 4$, & $\lambda = 0$ & & & \\[-0.01ex]
     $\alpha$ & $\phi $ &  $\phi_f$ & N & $n_s$ & r & $n_T$ \\
     \hline
     0 & 18.107 & 2.87070 & 40 & 0.92680 & 0.39041 & -0.0488\\
     0.01 & 25 & 2.85169  & 80  & 0.9647 & 0.18 & -0.0225 \\
     0.05 & 21.5 & 2.77846 & 70 & 0.9642 & 0.14807 & -0.0185\\
     0.1 & 19 & 2.69285   & 67  & 0.965622 & 0.097731 & -0.0122\\
     \hline
     Potential, & $V=V_0\frac{\alpha\phi^2}{1+\alpha\phi^2}$, &  & $\lambda = 0$ & & & \\[-0.01ex]
     $\alpha$ & $\phi $ &  $\phi_f$ & N & $n_s$ & r & $n_T$ \\
     \hline
     1 & 4.2  & 0.834039 & 43  & 0.964157 & 0.00522 & -0.00065\\
     2 & 3.55 & 0.707107 & 42 & 0.964126 & 0.003698 & -0.00046 \\
     %&  &  &  &  &  & \\ 
     \hline
 \end{tabular}
 \caption{\label{table:4} The e-fold number $N$ and the spectral index parameters $n_s$, $r$ and $n_T$ are presented here corresponding to $\lambda=0$ for all three potentials.}
 \end{table}
%%%
Note that in the Einstein gravity theory, the predicted value of $r$ and $n_T$ are slightly different than than of the modified gravity theory for the same set of potential parameter choice. From Table.~(\ref{table:2}) and Table.~\ref{table:4}), we have noticed that for $p=4, \alpha=0.01$ value, although $r$ value is little less in $\lambda=0$ case but $N$ is beyond $40-70$ whereas in $\lambda\neq 0$ case $N$ resides between $40-70$. 
It is also eminent that  out of the three potentials, the potential $V=V_0\frac{\alpha\phi^2}{1+\alpha\phi^2}$ gives the best results for e-fold and CMB parameters such as $n_s,~r$ and $n_T$ for  both $\lambda = 0$ as well as  $\lambda \neq 0$.
\newpage
\section{Analysis and Conclusion:}
\begin{figure}[htb]
 \centering
  \includegraphics[width=8.5cm]{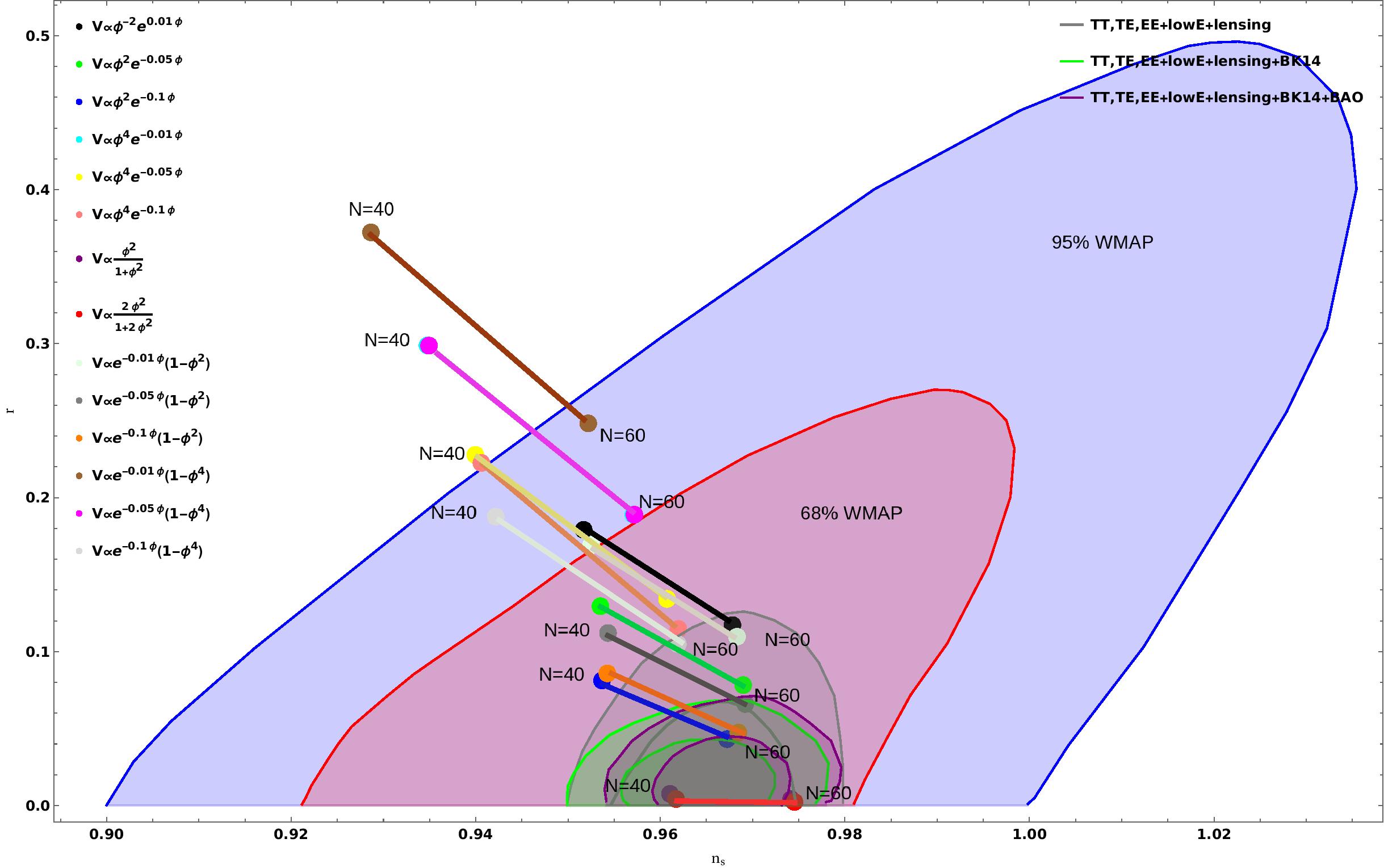}
  \caption{ \label{Plot1}{(Color online) Constraints on $n_s$ and $r$ from CMB measurements of different potential. Shaded regions are allowed by WMAP measuremnts, PLANCK alone, PLANCK+BK15, PLANCK+BK15+BAO upto $68\%$ and $95\%$ Confidence Level.}}
\end{figure}

In this paper, we have gone through the basics of slow-roll inflation in the context of modified gravity approach. We have analyzed inflationary cosmology for a particular form of $f(R,T) = R + 2 \lambda T$. This option has been extensively researched in the literature and is typically offered as an alternate strategy to deal with various cosmological issues, such as dark energy and dark matter. A discernible change to the results can also be produced by changing the functional form of $f(R,T)$ and its analysis is outside the purview of this paper. In this manuscript, we have focused on three different potentials$-$ $\phi^p e^{-\alpha\phi}$, $(1-\phi^p)e^{-\alpha\phi}$, $\frac{\alpha\phi^2}{1+\alpha\phi^2}$ to study the inflationary scenario in two context, one by taking the modified gravity $\lambda$ into account and the other by switching off $\lambda$ i.e. in normal Einstein gravity. We have calculated the slow-roll parameters, e-fold number and spectral index parameters in the case of a scalar field ($\phi$) minimally coupled to modified gravity. In order to do so, we have calculated $n_s$, $r$, $n_T$, $N$ for a particular value of field $\phi$ and have taken a range of $\lambda$ values which match with the spectral index data given by PLANCK2018 and WMAP. \\
In Fig. (\ref{Plot1}), we have shown the variation of $n_s$ and $r$ for all three potentials of inflationary expansion. 
%%%
The blue and red shaded region corresponds to WMAP data upto $95\%$ and $68\%$ C.L whereas grey, green and purple shaded regions corresponds to PLANCK, PLANCK+BK15, PLANCK+BK15+BAO respectively. The $N=40$ and $60$ values correspond to potentials $V=V_0\phi^pe^{-\alpha\phi}$ for i) $p=2$, $\alpha=0.01$, $0.05$, $0.1$, ii) $p=4$, $\alpha=0.01$, $0.05$, $0.1$ which are represented by black, green blue, cyan, yellow and peach lines and for potential $V=V_0(1-\phi^p)e^{-\alpha\phi}$ with i) $p=2$, $\alpha=0.01$, $0.05$, $0.1$, ii) $p=4$, $\alpha=0.01$, $0.05$, $0.1$, are shown in white, grey, orange, brown, magenta and light grey respectively. Purple and red lines are for potential $V=V_0\frac{\alpha\phi^2}{1+\alpha\phi^2}$ with $\alpha=1$, $2$. From Fig. (\ref{Plot1}), we can say that all the potentials are within WMAP data (at least $N=60$) whereas $V=V_0\frac{\alpha\phi^2}{1+\alpha\phi^2}$ is in the limit of PLANCK+BK15+BAO.
From Table \ref{table:4}, we have noticed that for a fixed $p$, $\phi$ value decreases with increasing $\alpha$ along with decreasing $N$. On the other hand, from Table \ref{table:1} and \ref{table:2}, we can see for $\lambda \neq 0$, we are getting good results for $p=2$, $\alpha=0.1$, $0.1$. As an example, for $V=V_0\phi^pe^{-\alpha\phi}$ and $p=2$, $\alpha=0.01$ we obtained $r=0.131321$, $0.128031$ for $\lambda=0$, $-0.114110$ respectively. So, $r$ value is little less for $\lambda\neq 0$ although $n_s$ is lying within $3\sigma$ limit of C.V. Similarly for other potentials, we can compare the cosmological parameters value from Table \ref{table:4} and Table \ref{table:1}, \ref{table:2}, \ref{table:3}.\\
 We infer that $V=V_0\frac{\alpha\phi^2}{1+\alpha\phi^2}$ fits best for all cosmological parameters with observational data given by PLANCK2018 and WMAP. The constraints on the potential parameters are found to satisfy the desired number of e-fold of inflationary expansion i.e. $40<N<60$ for a range of modified gravity parameter $\lambda$ for each potential. Finally, we can say by taking modified gravity into account, the predictions to tensor-to-scalar ratio and the tensor spectral index can be improved for the given set of potentials discussed in this paper. In normal Einstein gravity, we find $r$ and $n_T$ slightly higher than what we found in modified gravity.
\section{Acknowledgment}
AR would like to thank BITS Pilani K K Birla Goa campus for the fellowship support. PS would like to thank Department of Science and Technology, Government of India for INSPIRE fellowship.  PKD would like to thank Kinjal Banerjee for useful discussion.

\end{document}